\documentclass[useAMS, usenatbib]{mn2e}

\usepackage{aasmacros}
\usepackage{amssymb} 
\usepackage{subcaption} 
\usepackage{booktabs} 
\usepackage{multirow} 
\usepackage{amsfonts}
\usepackage{lmodern}
\usepackage{amsmath}
\usepackage{float}
\usepackage{graphicx}
\usepackage{soul}
\usepackage{color}

\title{Fast cold gas in hot AGN outflows} \author[Costa et al.]{Tiago Costa\footnotemark[1], Debora
  Sijacki and Martin G. Haehnelt \\
  Institute of Astronomy and Kavli Institute for Cosmology,
  University of Cambridge, Madingley Road, Cambridge CB$3$ $0$HA, UK}

\begin{document}
 
\maketitle

\begin{abstract}
Observations of the emission from spatially extended cold gas around
bright high-redshift QSOs reveal surprisingly large velocity widths exceeding $2000 \, \rm km \, s^{-1}$, out to 
projected distances as large as $30 \, \rm kpc$. The high velocity widths have been interpreted as the 
signature of powerful AGN-driven outflows. Naively, these findings appear in tension with hydrodynamic models in which
AGN-driven outflows are energy-driven and thus very hot with typical
temperatures $T \gtrsim 10^{6 \-- 7} \, \rm K$.  Using the moving-mesh code
{\sc Arepo}, we perform `zoom-in' cosmological simulations of a $z \sim 6$
QSO and its environment, following black hole growth and feedback via
energy-driven outflows. 
In the simulations, the QSO host galaxy is surrounded
by a clumpy circum-galactic medium pre-enriched with metals  due to
supernovae-driven galactic outflows. As a result, part of the AGN-driven hot
outflowing gas can cool radiatively, leading to large  amounts ($\gtrsim 10^9
\, \rm M_\odot$) of cold gas comoving with the hot bipolar outflow. This
results in velocity widths of spatially extended cold gas similar to those
observed. 
We caution, however, that gas inflows, random motions in the deep
potential well of the QSO host galaxy and cooling of supernovae-driven winds contribute
significantly to the large velocity width of the cold gas in the simulations,
complicating the interpretation of observational data. 

\end{abstract}


\section{Introduction}
\renewcommand{\thefootnote}{\fnsymbol{footnote}}
\footnotetext[1]{E-mail: taf34@ast.cam.ac.uk}

Bright quasars (QSOs) with luminosities $L \,\gtrsim\, 10^{47} \, \rm erg \,
s^{-1}$ powered by supermassive black holes with masses $\gtrsim 10^9 \, \rm
M_\odot$ exist already at $z \,=\, 6 \-- 7$ \citep{Fan:01, Willott:07, Mortlock:11, DeRosa:14}.  Their extreme luminosities
are thought to be generated via gravitational accretion of baryonic material
onto the black hole \citep{Lynden-Bell:69, Rees:84}.  It has been repeatedly
argued that it only takes a small fraction of the emitted energy to
couple with its host galaxy to unbind it entirely \citep{Silk:98,
  Haehnelt:98, King:03, Fabian:12}.  Such feedback from AGN has been
identified as an important process in the evolution of galaxies due to its
promise in facilitating the observed rapid quenching of star formation in
massive galaxies \citep{Scannapieco:04, Springel:05c, Croton:06, Bower:06,  Dubois:13, Sijacki:14} .

Energy and/or momentum deposition by AGN is certainly capable of launching
powerful outflows extending to galactic scales of tens of kpc in high-redshift
QSOs \citep[e.g.][]{Costa:14a, Costa:14b}.  Whether the coupling of energy
and/or momentum to the ISM occurs (magneto-)hydrodynamically via the shocking
of a fast inner wind or via direct radiation pressure is, however, intensely debated.  
The inferred momentum flux in observed outflows often appears to exceed the momentum flux $L/c$  of the AGN radiation field \citep[e.g.][]{Sturm:11, Cicone:14a, Genzel:14}. Here, we assume that the wind is driven hydrodynamically. In this picture, the observed momentum fluxes $> L/c$ favour an `energy-driven' wind over a `momentum-driven' model \citep[][]{Zubovas:12, Faucher-Giguere:12, Nayakshin:14, Costa:14b}. 

\begin{figure*}
\centering \includegraphics[scale = 0.45]{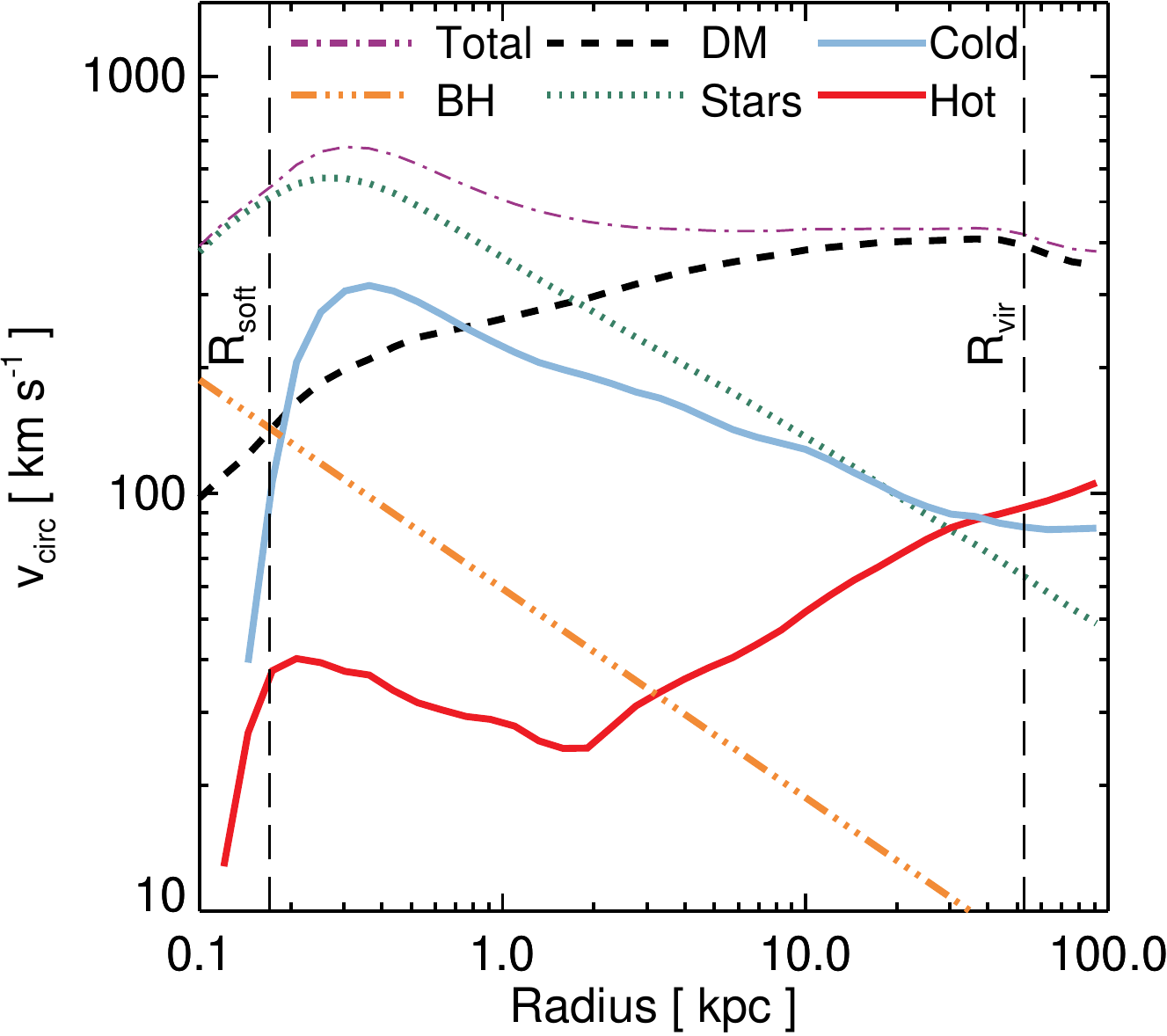}
\centering \includegraphics[scale = 0.44]{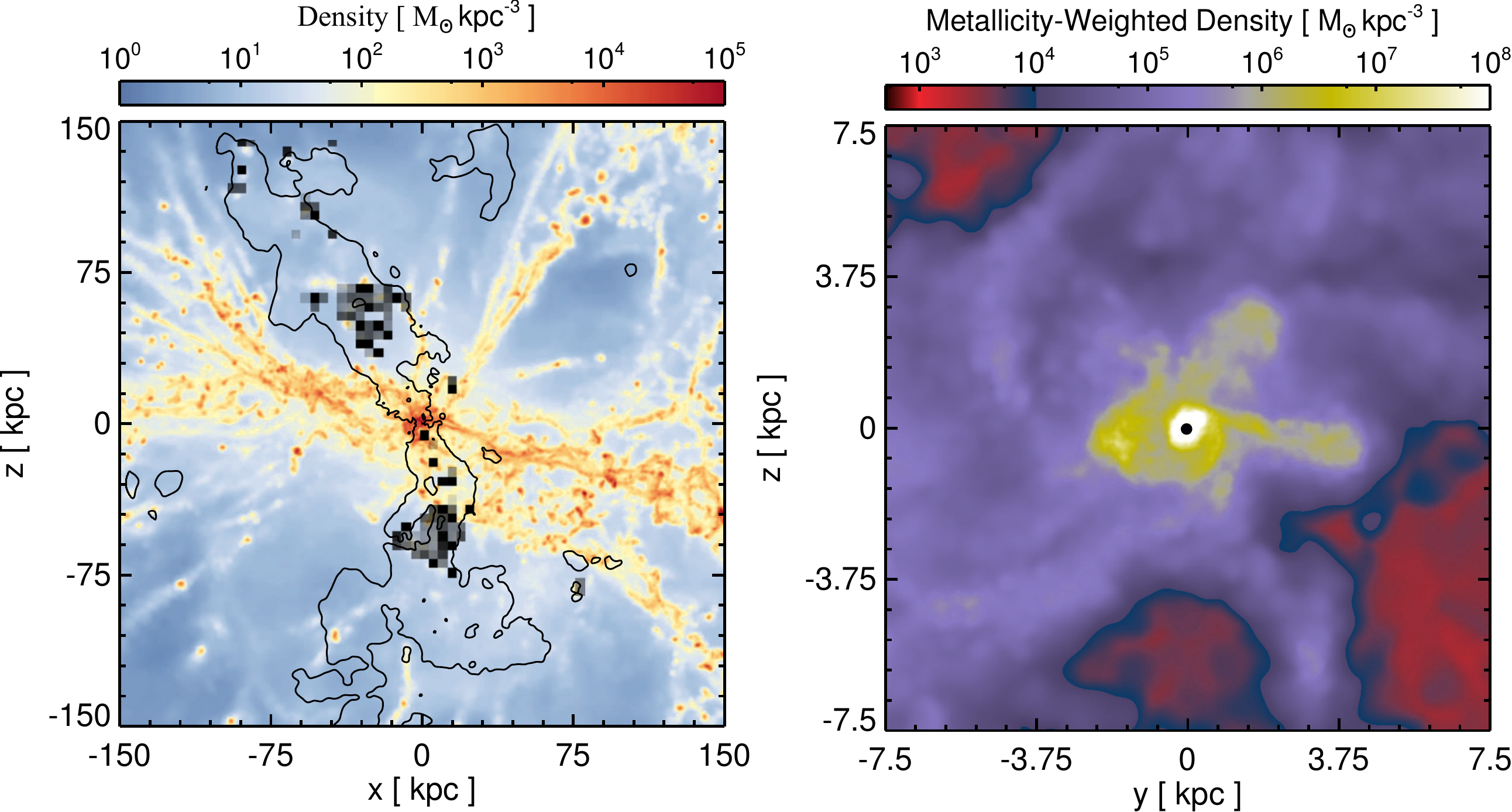}
\caption{{\bf Left:} Circular velocity ($v_{\rm circ} = \sqrt{GM(<r)/r}$) of dark matter, hot and cold gas,
  stars and the central black hole of the QSO host halo in the simulation
  with both supernovae and AGN feedback at $z \,=\, 6.6$. Vertical lines show the location of the gas softening length ($R_{\rm soft}$) and the virial radius of the host halo ($R_{\rm vir}$). {\bf Middle:} Gas
  density projected along a slab with thickness $300 \, \rm kpc$ at $z \,=\,
  6.6$. The contour shows the hot component of the outflow with a velocity of $500 \, \rm km
  \, s^{-1}$. The location and radial velocity-weighted mass of outflowing
  cold clouds is marked with pixels, with darker colours indicating a higher mass in the cold phase. {\bf Right:}  Zoom-in into the
  central $15 \, \rm kpc$ of the QSO host galaxy showing metallicity-weighted gas
  density. The remarkably compact central disc is oriented perpendicularly to the
  large-scale outflow and is fed by a system of cold gas filaments. The location of the most massive black hole ($M_{\rm BH} \,=\, 8.2 \times 10^8 \, \rm M_\odot$) is marked with a black circle.}
\label{outflow_project}
\end{figure*}

One of the potential challenges to energy-driven outflow models is their
prediction that the outflow is not only fast, but also hot and tenuous.
Observations, however, suggest that a significant component of AGN-driven
outflows is cold and dense molecular gas \citep[e.g.][]{Aalto:12, Cicone:14a}. 
It has been argued that, in the absence of magnetic fields, the hot outflow is unable to fully drag cold gas by ram pressure without disrupting it \citep[see e.g.][]{McCourt:14}. A possible route to a cold component in hot energy-driven outflows is instead radiative cooling of the shocked swept-up material \citep{Zubovas:14a}.  In this Letter, we
investigate whether the cooling of hot energy-driven outflows in
cosmological simulations of a bright QSO at $z \,\sim \, 6$ can account for 
a cold outflowing component.  We compare our predictions for
the spatial distribution and kinematics of the cold gas with the recent interferometric PdBI observations of an AGN-driven cold outflow detected in a luminous $z \approx
6.4$ QSO \citep{Cicone:14b}.

\section{Numerical Simulations}

We employ the moving-mesh code {\sc Arepo} \citep{Springel:10} in order to
follow the hydrodynamics of the cosmic gas component, treated as an ideal
inviscid fluid. We include radiative primordial cooling as modelled in \citet{Katz:96} combined with metal-line cooling as treated in \citet{Vogelsberger:13}.
We also include heating due to a spatially uniform time-dependent UV background \citep{Faucher-Giguere:09}.
Our simulation suite consists of `zoom-in'
resimulations of the six most massive haloes found in the Millennium volume
\citep{SpringeL:05b} at $z \,=\, 6.2$ and captures the likely cosmological
sites of the brightest high redshift QSOs \citep{Sijacki:09, Costa:14a}\footnote{As discussed in \citet{Costa:14a}, the volume probed by the Millennium simulation is still a factor $\approx 100$ smaller than that probed by surveys of high-redshift QSOs. The host haloes should be even more massive than those considered in our simulations if the duty cycle is close to unity.}.  In
order to model energy-driven outflows, we adopt the prescription described in
\citet{Costa:14b}, which
successfully recovers the analytical wind solution 
for energy-driven shells. 
Note that we now follow black hole accretion self-consistently as in \citet{Costa:14a}.
We also investigate simulations with
supernovae-driven outflows (\citet{Springel:03}, see also \citet{Costa:14a}).
In this Letter, we isolate the effects of feedback by analysing four different
runs: 1. simulation without supernovae or AGN feedback (No Feedback),
2. simulation with supernovae-driven winds with mass loading $\eta \,=\, 1$ and 
velocity $1183 \, \rm km \, s^{-1}$ comparable with the escape speed from the AGN host halo (SN Feedback), 3. simulation with energy-driven AGN winds (AGN
Feedback), and 4. simulation with combined supernovae- and AGN-driven outflows (SN+AGN
Feedback). To distinguish different gas phases, we consider a
temperature cut of $\geq 10^6 \, \rm K$ for `hot gas' and $\leq 5 \times
10^4\, \rm K$ for `cold gas' additionally including `star-forming' gas lying
along the ISM effective equation of state\footnote{We take the `cold gas' in our simulations as proxy for molecular gas, which cannot be distinguished in our simulations due to unavoidable lack of resolution.}. Since most of the gas at
temperatures $\leq 5 \times 10^4 \, \rm K$ has cooled to the temperature floor
of $10^4 \, \rm K$, our results are only mildly sensitive to this
definition. All spatial coordinates are given in physical units. 

\section{Results}

\begin{figure*}

\includegraphics[scale = 0.39]{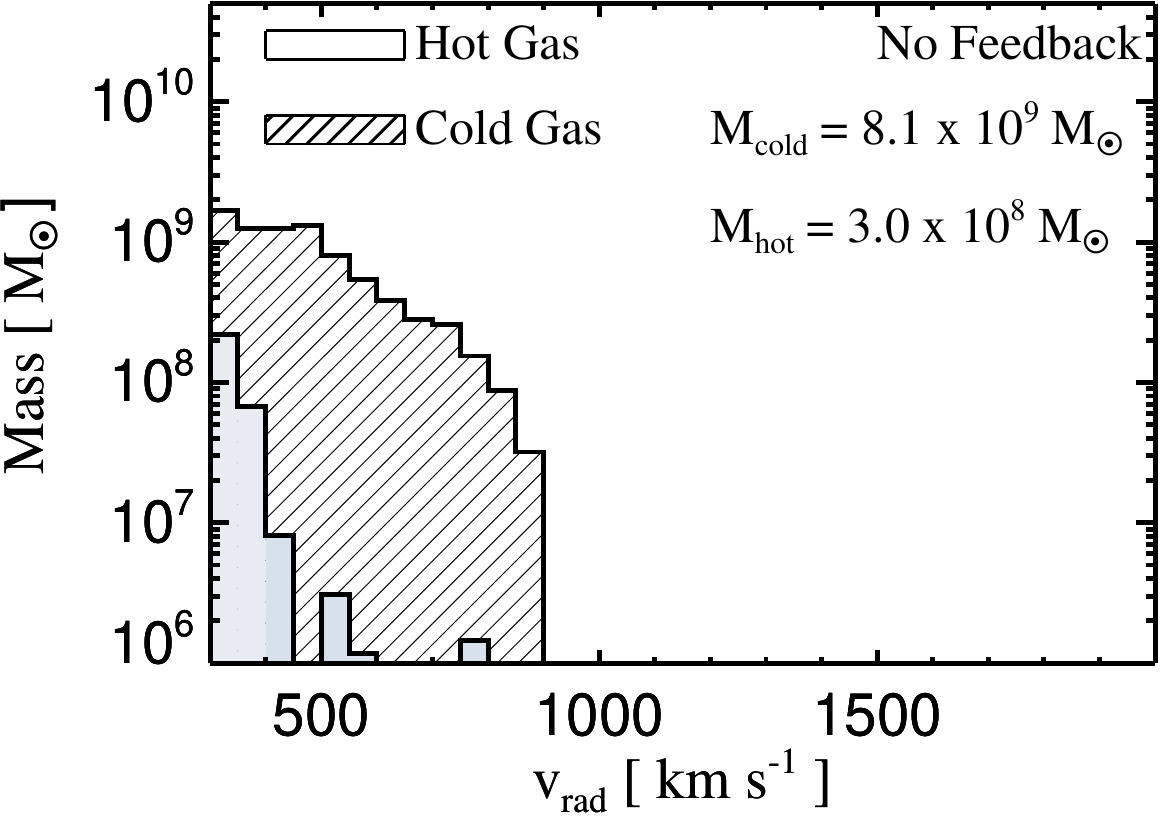}
\includegraphics[scale = 0.39]{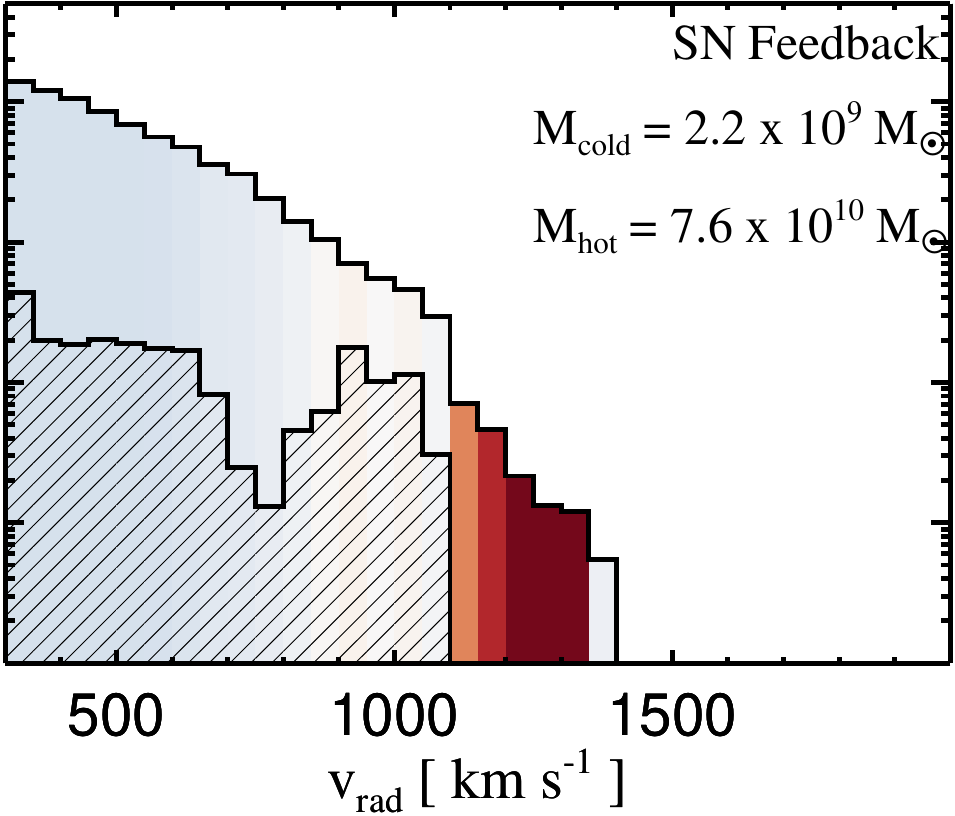}
\includegraphics[scale = 0.39]{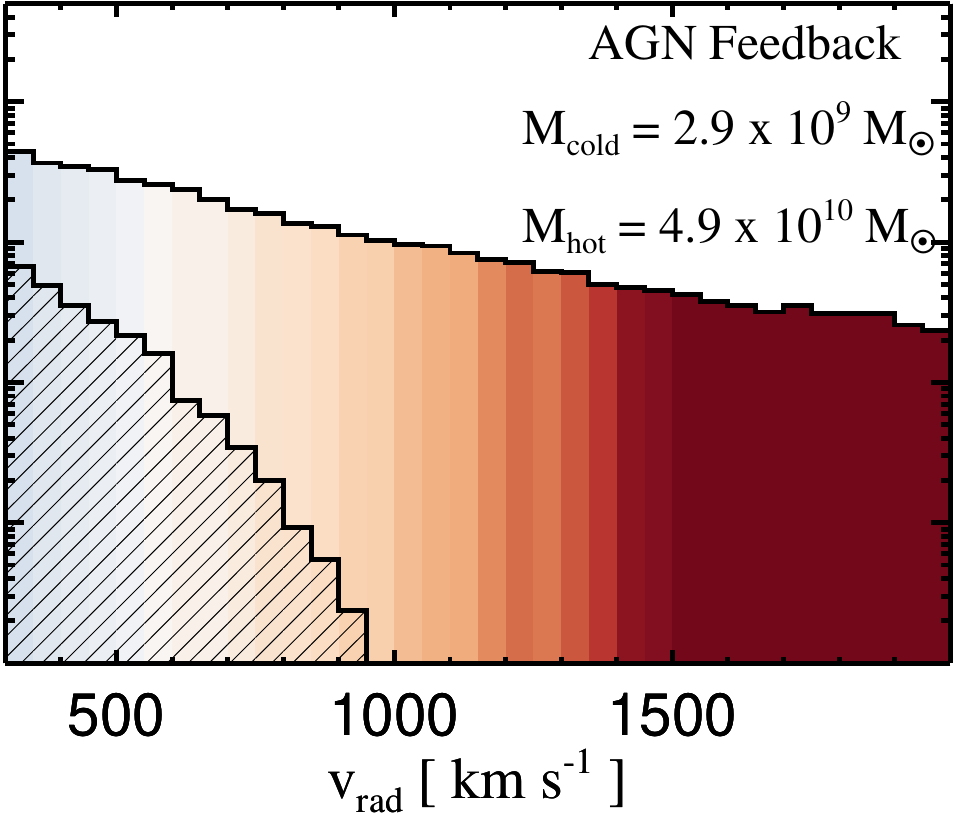}
\includegraphics[scale = 0.39]{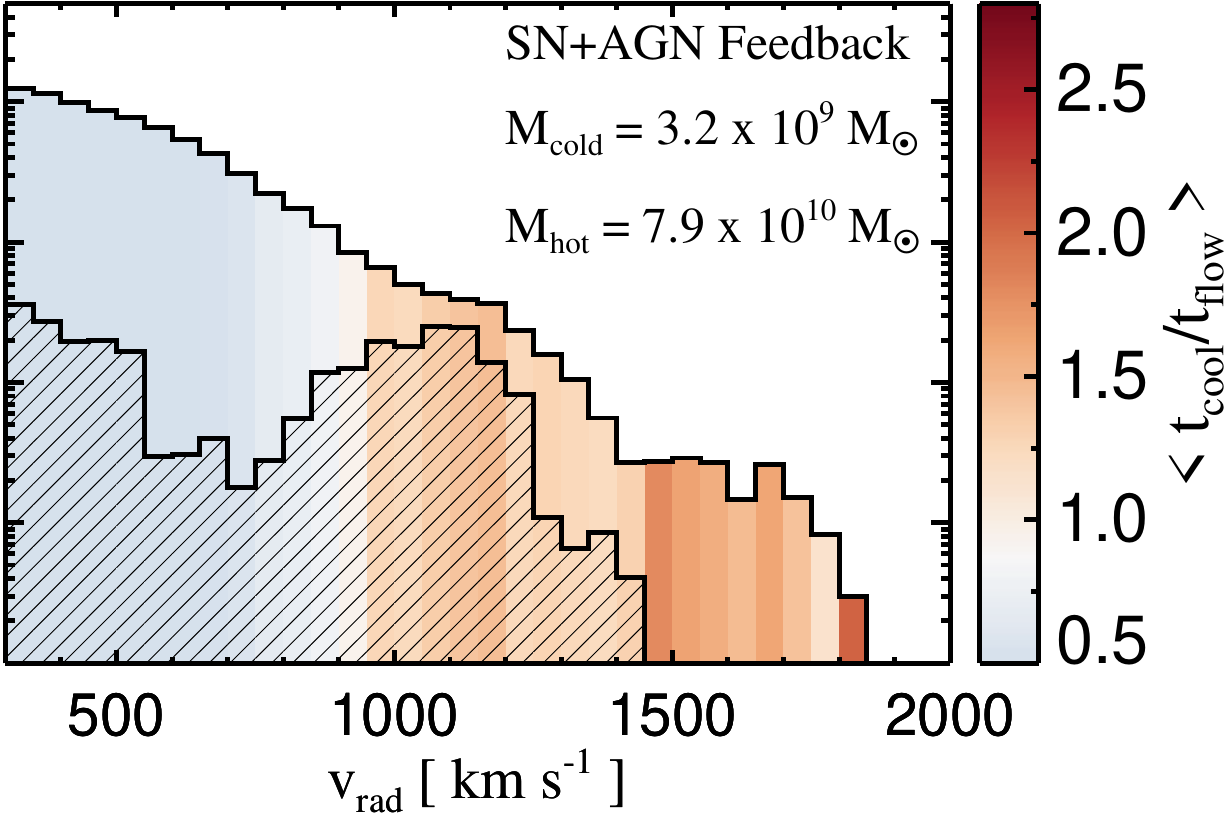}

\includegraphics[scale = 0.39]{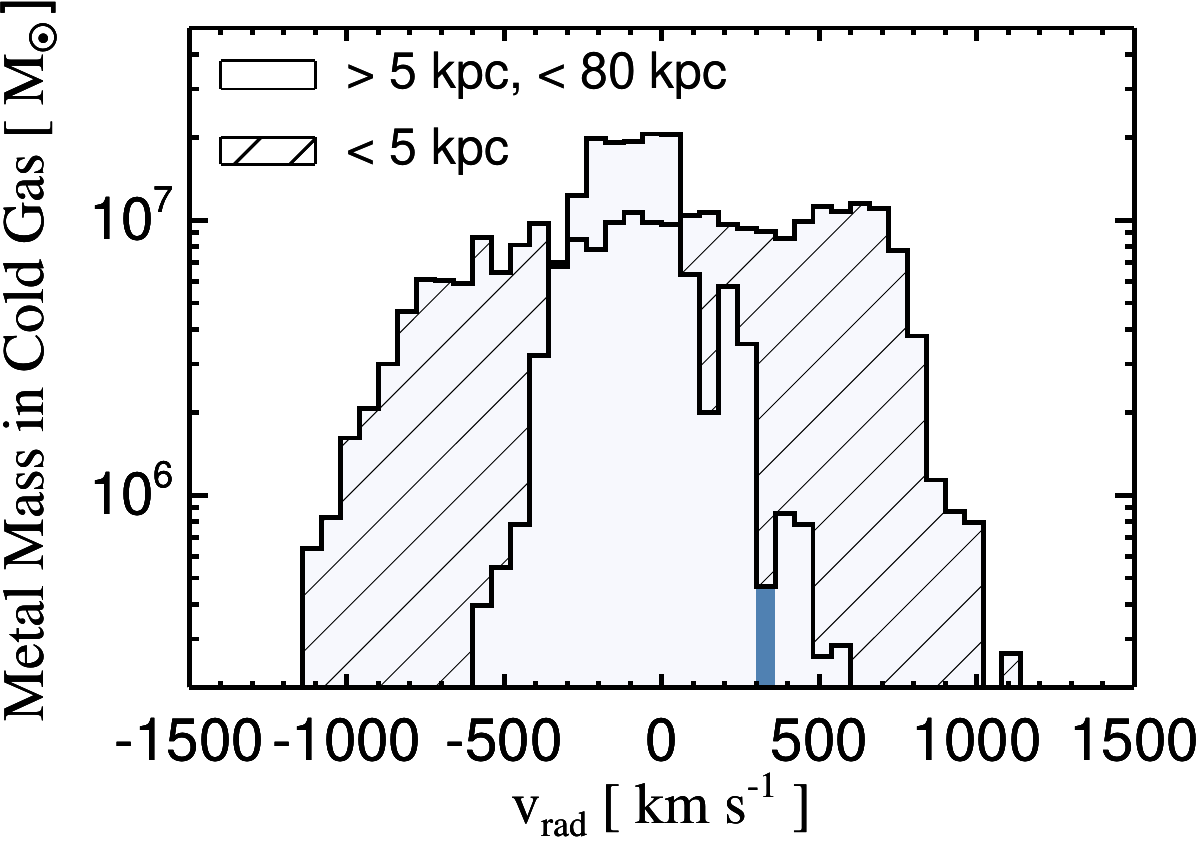}
\hspace{-0.3cm}
\includegraphics[scale = 0.39]{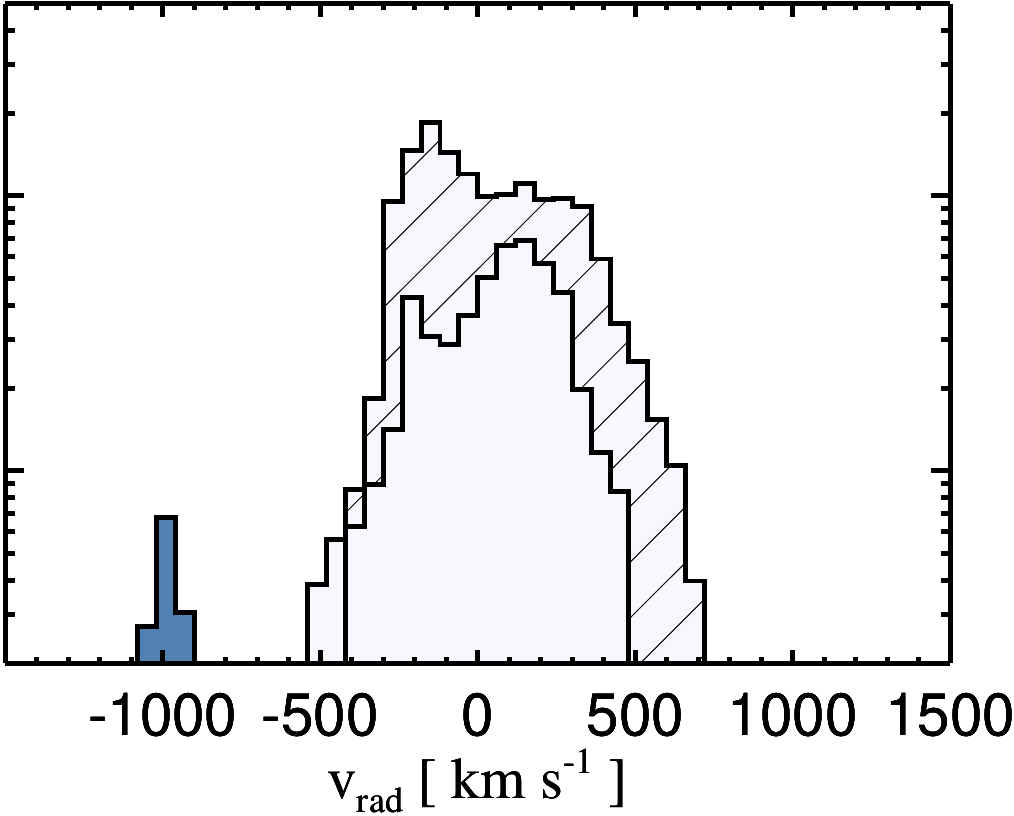}
\hspace{-0.3cm}
\includegraphics[scale = 0.39]{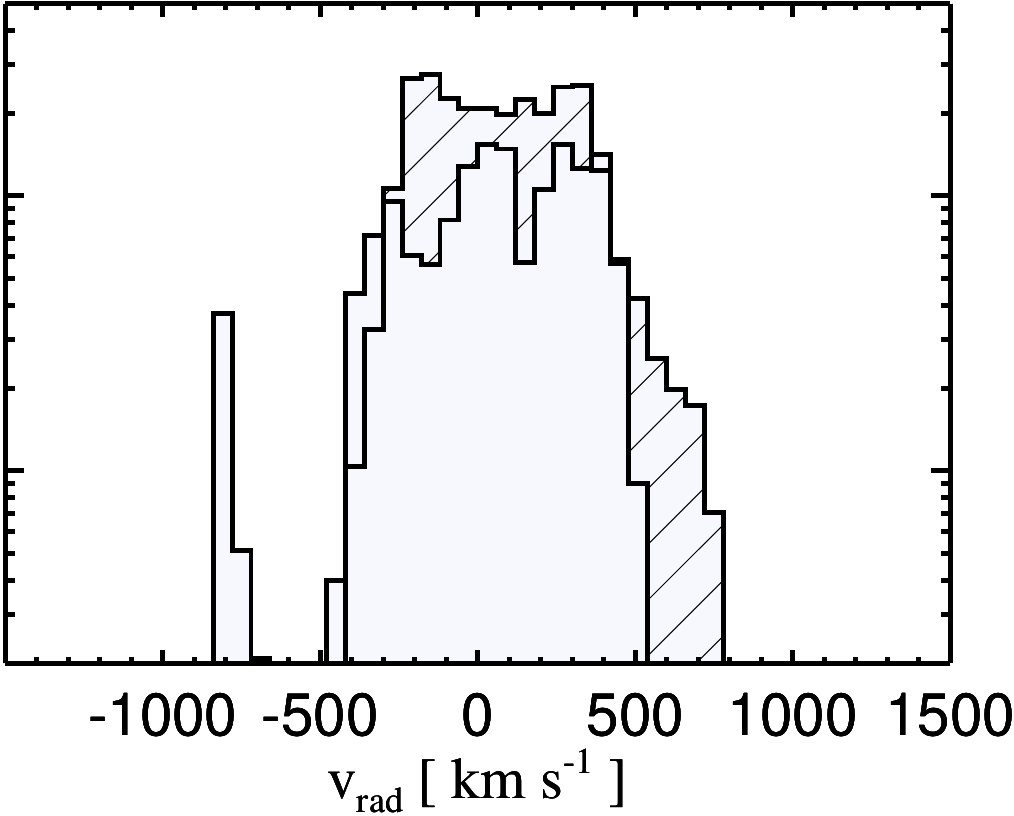}
\hspace{-0.3cm}
\includegraphics[scale = 0.39]{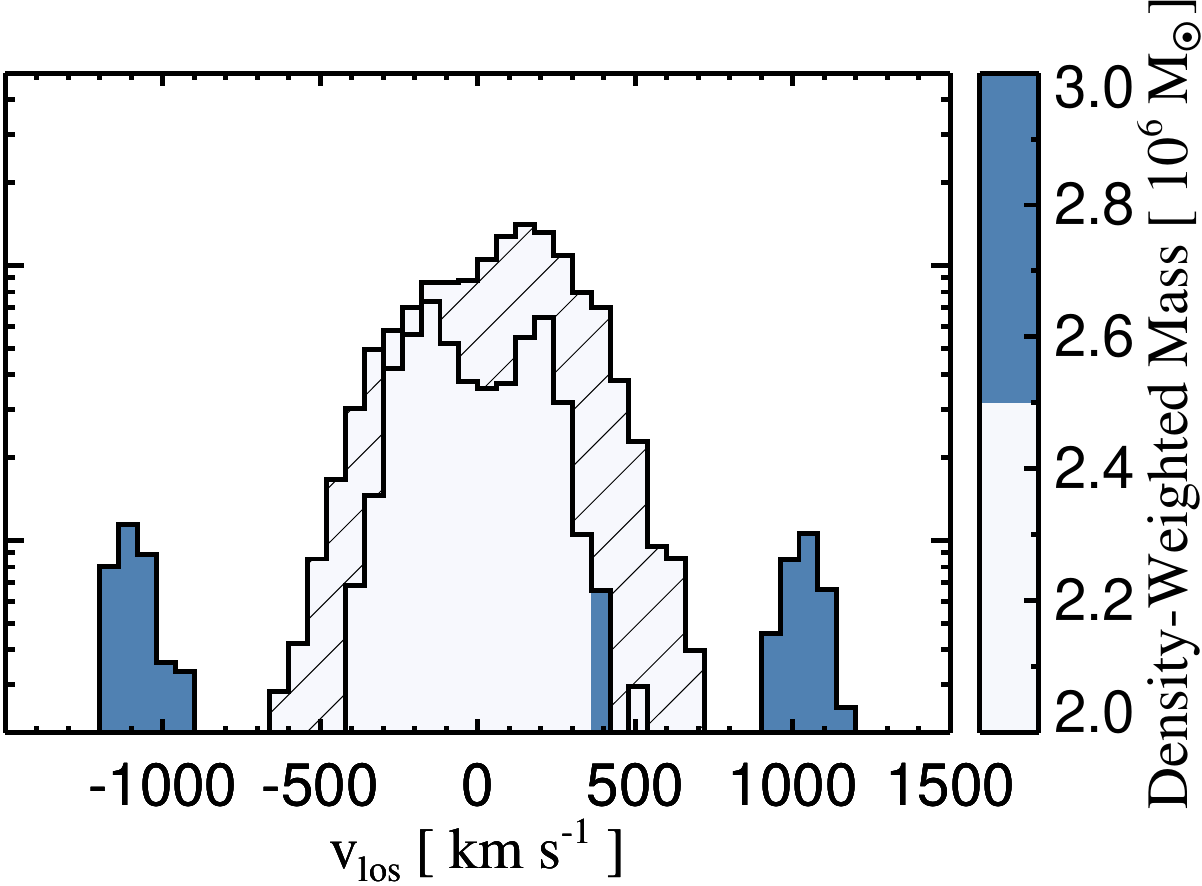}

\caption{{\bf Top row:} Cold gas mass per radial velocity bin within a sphere of radius $300 \, \rm kpc$ centred on the QSO host galaxy. Bins
  are colour-coded according to the average ratio of cooling to flow
  time for the hot component $\langle t_{\rm cool}/t_{\rm flow} \rangle$. Cold comoving gas forms whenever the flow time becomes comparable to the cooling
  time. This occurs for gas of speeds of up to $1500 \, \rm km \, s^{-1}$ in simulation SN+AGN Feedback. A
  large amount of gas travels at faster speeds but does not cool sufficiently
  quickly and remains hot. The mass in cold and hot gas flowing out at speeds higher than $300 \, \rm km \, s^{-1}$ is labelled in each plot. Note that due to different black hole growth histories in simulations with different feedback, an AGN-driven outflow shown here only occurs at a slightly lower redshift of $6.3$ in simulation AGN Feedback. {\bf Bottom row:} Metal-enriched cold gas mass per
  line-of-sight velocity bin shown within radii $5 \,
  \rm kpc$ (shaded) and between $5$ and $80 \, \rm kpc$ (coloured) at $z \,=\, 6.6$ ($z \,=\, 6.3$ in the AGN Feedback run). 
  The line-of-sight was chosen to be perpendicular to the central gas disc. 
  Colours give the density-weighted average mass in each bin. High
  gas speeds occur without any feedback, but all the fast moving gas is
  concentrated in the central regions of the halo. In order to produce
  spatially extended high velocity cold gas, strong outflows are
  required.}
\label{spectra}
\end{figure*}

At $z \,=\, 6.6$, the simulated QSO has a bolometric luminosity of $L_{\rm
  bol} \,\approx\,10^{47} \, \rm erg \, s^{-1}$ powered by a
black hole of mass $M_{\rm BH} \,=\, 8.2 \times 10^8 \, \rm M_\odot$ and hence has
an Eddington ratio $\lambda_{\rm Edd} \,=\, \dot{M}_{\rm BH}/\dot{M}_{\rm Edd}
\,\approx\, 1$, all in line with observed properties of $z \gtrsim 6$ QSOs
\citep[e.g.][]{DeRosa:14}.  In our simulations, the bright QSO is located in a
very massive halo with $M_{\rm halo} \,\approx\, 3 \times 10^{12} \, \rm
M_\odot$ within a tightly bound galaxy with a gaseous disc and a very compact flattened
stellar bulge. The galaxy has a stellar mass of $2.2 \times 10^{10} \, \rm M_\odot$ within a radius $\approx 0.3 \, \rm kpc$ and is undergoing a starburst with a total star formation rate of $\approx 470 \, \rm M_\odot \,
yr^{-1}$. Note that star formation rates inferred from (sub)millimeter observations of
high redshift QSO hosts can be even higher, reaching up to $3000 \, \rm
M_\odot \, yr^{-1}$ \citep[e.g.][]{Bertoldi:03}.  The QSO host halo
has a circular velocity $v_{\rm circ} \approx 400 \, \rm km \,s^{-1}$ at the
virial radius $R_{\rm vir} \approx 53 \, \rm kpc$ (see Fig.~\ref{outflow_project}) and is continuosly fed by three massive cold filaments, leading to the formation of the central bulge with a maximum circular velocity $v_{\rm circ} \approx 700 \, \rm
km \,s^{-1}$.

As previously reported, energy and momentum injection in the
QSO host galaxy via star formation and black hole accretion leads to
powerful galactic outflows in our simulations
\citep{Costa:14a}. 
We here investigate whether the initially hot outflows are able to cool radiatively.
In order for the outflow to cool, the flow time $r/v_{\rm rad}$ must exceed the cooling time $u_{\rm int}/ \left( n_{\rm H}^2 \Lambda \right)$ of
the shocked swept-up material. The histograms displayed in the top row of Fig.~\ref{spectra}
show the distribution of radial velocities for hot (coloured) and cold
(line-filled) gas for our four simulations.  
Due to the depth of the central potential well, positive radial velocities of up to $\approx 900 \, \rm km
\,s^{-1}$ occur without either AGN or supernovae feedback, while the
distribution of radially outflowing hot gas extends to $\approx 800 \, \rm km
\, s^{-1}$. The subsequent panels in the top row show that a hot outflowing
wind with velocities up to $\approx 1400 \, \rm km \, s^{-1}$ is launched in
the simulation with supernovae feedback only, while the simulation with AGN
feedback produces much faster hot outflows with velocities well beyond $2000\, \rm km \, s^{-1}$. 

The histograms are colour-coded according to the bin-averaged ratio of cooling
time to flow time $\langle t_{\rm cool}/t_{\rm flow} \rangle$ of the hot component.  
The red tail of the histogram for gas with velocities $>1500 \, \rm km \, s^{-1}$ in the simulations with AGN feedback and with both AGN and supernovae feedback
represents gas with $\langle t_{\rm cool}/t_{\rm flow}\rangle \approx 1 \-- 3$.  This
ultrafast component of the hot gas cannot cool in the outflow. 
However, for somewhat slower gas, $\langle t_{\rm cool}/t_{\rm flow}\rangle \approx 1$ 
despite the still rather high speeds  ($v_{\rm rad} \lesssim 1300 \, \rm km \, s^{-1}$).    
In the simulation with supernovae feedback, a large fraction
of the hot outflow can actually cool, leading to a significant enhancement of
the amount of cold gas with radial velocities in the range $500 \-- 1100 \,
\rm km \, s^{-1}$. 
It is, however, the combined effect of supernovae and AGN feedback that leads to the largest mass
of fast ($\gtrsim 1000 \, \rm km \, s^{-1}$) outflowing cold gas and the highest outflow speeds ($\approx 1400 \, \rm km \, s^{-1}$). 
Note that in all simulations, the total mass of cold gas flowing out with speeds $v_{\rm rad} > 300 \, \rm km \, s^{-1}$ exceeds $10^{9} \, \rm M_\odot$. 
Note that we verified that additional Compton heating from the QSO does not produce a significant effect over a flow time at radii of tens of kpc \citep[e.g.][]{Costa:14b}.

In the bottom row of Fig.~\ref{spectra}, we show the mass of metal-enriched cold gas as a function of 1D (line-of-sight) velocity along the
pole of the QSO host galaxy. 
In the No Feedback simulation, all of the rapidly moving cold
gas is located within the innermost $5 \, \rm kpc$ of the halo.
 The potential well of the QSO host is so deep (see
 Fig.~\ref{outflow_project}) that gravitational motions alone produce a
 velocity width for the cold gas\footnote{Note that in the absence of feedback processes, gas may become unrealistically compact.} of about $2000\, \rm km \, \rm s^{-1}$ within a radius of $5 \, \rm kpc$. 
In simulations SN Feedback and AGN Feedback, the outflows clear out part of the gas in the
central region of the galaxy, thus reducing the total mass in cold gas.
The velocity widths of the cold gas are also somewhat smaller. 

\begin{figure*}
\centering \includegraphics[scale = 0.63]{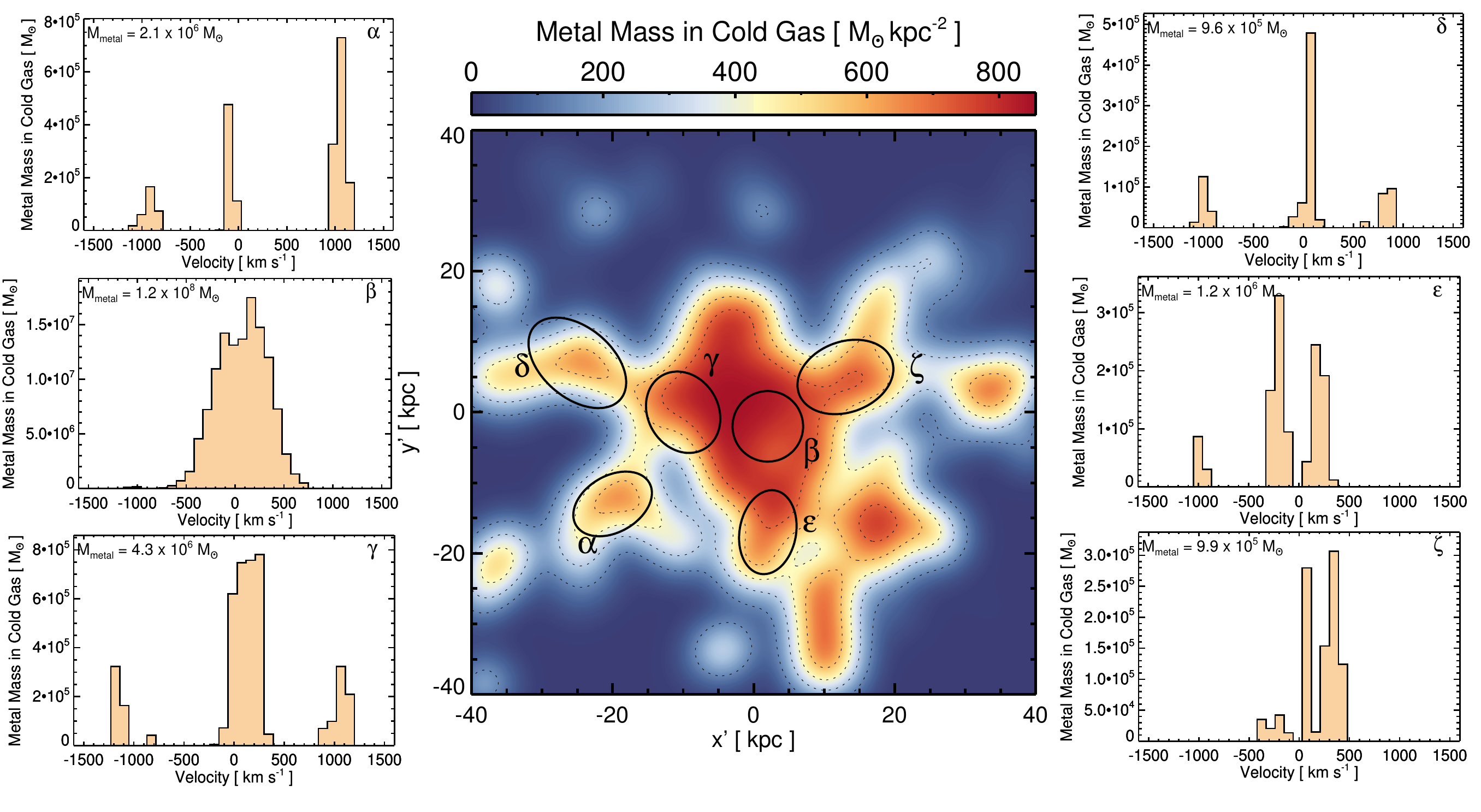}
\caption{Cold gas mass in metals projected along the simulation volume and
  convolved with a Gaussian filter with FWHM of $6.7 \, \rm kpc$ to
  match the resolution of observations by \citet{Cicone:14b}. Cold
  outflowing gas is spatially extended out to (projected) distances of  $20 \-- 30 \, \rm kpc$ and
  is characterised by a clumpy and very irregular morphology. The kinematics
  of the outflowing cold gas is illustrated by spectra extracted from various
  regions of the image. The individual spectra differ substantially from
  region to region and can be characterised by red and
  blue `wings' symmetric in velocity space, tracing gas flowing out at speeds up to $1200 \, \rm km \,
  s^{-1}$ or by a narrow component tracing low velocity gas. The narrow component is also spatially extended.}
\label{outflowstructure}
\end{figure*}

The combined effect of supernovae and AGN feedback has, however, a much larger effect on the spatial and velocity distribution of the cold gas. 
The supernovae-driven outflows lead to a more metal-enriched clumpy environment around the host galaxy
which drives much more efficient cooling within the AGN-driven fast hot
outflow. As a result, the velocity width of the cold gas between radii of $5$ and $80 \, \rm kpc$ increases to $\approx 2400 \, \rm km \,s^{-1}$.
The momentum flux of the cold gas embedded in the energy-driven outflow in our simulations is relatively low ($\approx 0.27 L/c$) comparable to the $1.00 \pm 0.14 L/c$ inferred by \citet{Cicone:14b}. The hot outflow driven by the AGN nevertheless contains a momentum flux of $\approx 3.3 L/c$.

The central panel of Fig.~\ref{outflow_project} pictures the hot and cold outflows launched in the simulation with both supernovae and AGN feedback. 
The hot outflow takes paths of least resistance \citep[see also][]{Zubovas:12,  Costa:14b} and propagates along the poles of the host galaxy (shown in the right panel), thereby avoiding regions of filamentary infall.  
The clumpier pockets of cold gas largely retain the geometry of the
hot outflow from which they cooled and are spatially extended with radial distances 
up to  $\approx 100 \, \rm kpc$ from the QSO.

Due to the anisotropy of the outflow, a spatial offset between
the blue and red spectral wings is expected for most outflow orientations.
For a bipolar outflow like that in our simulation, a spatial overlap between the
red and blue  wings of the velocity distribution requires  a rather 
special orientation  in which the line of sight is parallel to the symmetry axis of the outflow.  
The recent [CII] observations \citep{Maiolino:12, Cicone:14b} of the bright $z \,\approx\, 6.4$ QSO J1148+5152 show extremely broad wings extending to speeds of $\approx 1400
\, \rm km \, s^{-1}$ and also reveal widespread overlap between
the red and blue high velocity tails of the emitting gas distribution. Given the potential observational bias of preferentially selecting high redshift
QSOs residing in galaxies with approximately face-on orientations
\citep{Ho:07, Narayanan:08}, it appears therefore not unlikely that the outflow in J1148+5152
is indeed  bipolar and propagating along the line-of-sight.
Note that this means that the actual distance to the 
spatially extended gas is larger by probably about a least a factor two than the observed projected distances.
Note further, that out of our six resimulations, we have picked the one  with the most symmetric bipolar outflow. 
In most simulations in our sample, the outflows show complex geometries 
resulting in asymmetric or unipolar velocity distributions of the cold gas even for lines-of-sight along the main axis of the outflow.

In Fig.~\ref{outflowstructure}, we project the mass of cold metal-enriched
material within a slab of
thickness $500 \, \rm kpc$ along the outflow axis. The image was smoothed with
a Gaussian filter with FWHM of $6.7 \, \rm kpc$ to match the resolution
obtained by \citet{Cicone:14b}.  The distribution of cold outflowing material
strongly resembles the irregular morphology of the [CII] image of \citet{Cicone:14b} (see their Fig.~3). The projected spatial scale of $20
\-- 30 \, \rm kpc$ of the cold metal-enriched gas is also consistent with their reported value of $\approx 30 \, \rm kpc$.  
As in Fig.~3 of \citet{Cicone:14b}, we selected six regions in the QSO vicinity
and extracted `spectra' showing the mass of metal-enriched cold gas as a function of line-of-sight velocity.  Fast red- and blue-shifted outflowing cold gas is traced by symmetric features in velocity space, e.g. regions $\alpha$, $\gamma$, $\delta$, $\epsilon$, $\zeta$, existing throughout the image over spatial scales of a few tens of kpc.
Some of the regions (e.g. $\beta$) are instead mainly characterised by a narrow
component tracing `quiescent' cold gas, with a velocity width that varies from region to region.  We emphasise that as in \citet{Cicone:14b},
such a narrow component is also spatially extended over a scale of $\approx 30 \, \rm kpc$, but its velocity width is somewhat broader in our simulations.  
The spectra  in Fig.~\ref{spectra} show that  spatially extended `quiescent' gas is a common feature of every simulation, 
including that with no feedback. 
The narrow component traces gas assembling in the QSO host halo, moving at
large angles with respect to the line-of-sight. Part of this gas is in the
filaments feeding the halo and part is outflowing (in the simulations with
feedback). Unfortunately, the moderate signal-to-noise of \citet{Cicone:14b} precludes us from concluding 
whether the more continuous appearance of the observed spectra is due to observational noise or due to a 
more complex velocity distribution of cold material than captured by our simulations. 

\section{Discussion \& Conclusions}

Observations of the spatial distribution and kinematics of cold
gas surrounding bright high-redshift QSOs provide a remarkable opportunity to
probe at the same time the environment of these enigmatic objects and the
physical mechanism responsible for AGN feedback.  Our simulations show that
large velocity widths by themselves are not necessarily an indicator of strong
outflows.   The  rather massive  haloes expected to host these bright QSOs
develop  sufficiently deep potential wells that gravitational motions alone 
can result in velocity widths  as large as $\approx 2000 \, \rm km \,
s^{-1}$  for gas at the centre of the host galaxy. 

 Much more constraining are therefore the observations of
such velocity widths  in spatially extended gas reaching out to projected distances $\gtrsim 30 \, \rm kpc$ from the QSO. We have shown here
that radiative cooling of the hot shocked outflowing material can account for the formation of entrained cold gas in both supernovae-
and AGN-driven outflows at such distances. 
The large observed velocity at projected distances of $\gtrsim 30 \, \rm kpc$  \citep{Maiolino:12, Cicone:14b} however require a combination of
both.  In the absence of supernovae feedback, the AGN-driven outflows
that preferentially expand into the voids
encounter gas of too low density and metallicity for sufficient cooling
to take place. Supernovae-driven outflows pre-enrich 
the circum/inter-galactic medium
with metals and fill it with higher density clumps. If these are overrun by the fast hot AGN
outflow, the cooling time of the shocked hot gas significantly decreases, resulting in more vigorous cold and relatively dense outflows. The fastest outflow velocities ($\gtrsim 1400 \, \rm km \, s^{-1}$)
and highest outflow rates of cold gas therefore occur when supernovae- and
AGN-driven outflows work in tandem.
Strong further clues about the nature  of the cold outflows may thereby come from the 
astonishing blue/red symmetry of the velocity width distribution of the observed spatially extended gas in \citet{Cicone:14b}. In our simulations, this not only requires 
a rather special orientation of the line-of-sight, but also a highly symmetric (bipolar) outflow.
As we have shown, such highly symmetric outflows occur in our simulations, but this is by no means the generic configuration 
of the simulated outflows.  Characterising larger samples of spatially extended 
outflows of cold gas (as well as the strong selection effects of the observed samples) should thus provide 
valuable information on the environment of bright high-redshift QSOs as well as 
the physical mechanism driving the outflow.

\section{Acknowledgments}

We thank the referee for a helpful report and are grateful to Claudia Cicone, Roberto Maiolino and Sergei Nayakshin for many discussions and helpful suggestions.
All simulations used for this study were performed on the Darwin HPC Supercomputer based at the University of Cambridge, 
UK as part of the DiRAC supercomputing facility. TC acknowledges an STFC studentship and MGH and TC acknowledge
support by ERC ADVANCED Grant 320596 `The Emergence of Structure during the epoch of Reionization'.

\bibliographystyle{mn2e} 
\bibliography{references}

\end{document}